\documentclass[twocolumn,prb,groupedaddress,superscriptaddress,showpacs,showkeys]{revtex4}
\usepackage{epsfig}
\usepackage{graphicx}
\usepackage[utf8]{inputenc}
\usepackage{amsmath, amssymb}
\usepackage{braket}  
\usepackage{color}  

\newcommand{\Avr}[1]{\Braket{#1}}
\definecolor{amethyst}{rgb}{0.6, 0.4, 0.8}
\definecolor{bondiblue}{rgb}{0.0, 0.58, 0.71}

\begin{document}
 
\title{Influence of the nature of confinement on the melting of Wigner molecules in quantum dots}

\author{Dyuti~Bhattacharya}
\affiliation{Indian Institute of Science Education and Research Kolkata, Mohanpur Campus, India-741252}
\author{A.V.~Filinov}
\email{filinov@theo-physik.uni-kiel.de}
\affiliation{Christian-Albrechts-Universität zu Kiel, Institut für Theoretische Physik und Astrophysik, \\ Leibnizstrasse 15, D-24098 Kiel, Germany}
\affiliation{Joint Institute for High Temperatures RAS, Izhorskaya Str. 13, 125412 Moscow, Russia}
\author{Amit~Ghosal}
\email{ghosal@iiserkol.ac.in}
\affiliation{Indian Institute of Science Education and Research Kolkata, Mohanpur Campus, India-741252}
\author{M.~Bonitz}
\affiliation{Christian-Albrechts-Universität zu Kiel, Institut für Theoretische Physik und Astrophysik, \\ Leibnizstrasse 15, D-24098 Kiel, Germany}
\begin{abstract}
We analyze the quantum melting of two-dimensional Wigner molecules (WM) in confined geometries with distinct symmetries and compare it with corresponding thermal melting. Our findings unfold complementary mechanisms that drive the quantum and thermal crossovers in a WM and show that the symmetry of the confinement plays no significant role in determining the quantum crossover scale $n_X$. This is because the zero-point motion screens the boundary effects within short distances. The phase diagram as a function of thermal and quantum fluctuations determined from independent criteria is unique, and shows `melting' from the WM to both the classical and quantum ``liquids". An intriguing signature of weakening liquidity with increasing temperature, $T$, is found in the extreme quantum regime. The crossover is associated with production of defects. However, these defects appear to play distinct roles in driving the quantum and thermal `melting'. Our study will help comprehending melting in a variety of experimental traps - from quantum dots to complex plasma.

\end{abstract}



\maketitle

\section{Introduction} 
A condensed phase of matter, and its transition to a different one is sharply defined only in the thermodynamic limit~\cite{HuangBook87,YangLee52}. Yet, the mechanism driving the transition and the ``phases" can be discerned as a ``crossover"~\cite{Egger,Filinov01,GhosalNat07,Filinov08,Expt1stpara11} even in a confined geometry with finite number of particles. Such has been the scenario for the ``melting" of a Wigner molecule (WM)~\cite{Expt412} -- a puddle consisting of Coulomb-interacting particles (electrons). Under the influence of thermal or quantum fluctuations, a WM crosses over to a liquid-like phase. WM is a nano-scale version of two-dimensional (2D) Wigner crystal~\cite{2DWC34,Tanatar89}, an insulating phase of a 2D electron solid whose non-conducting nature arises from the propensity of the electrons to localize as far apart from each other as possible (consistent with their density) minimizing the free energy.

The study of the Wigner molecules found a large scientific quest primarily with the advances in nano-technology. They are easily tunable using electrostatic and magnetic methods~\cite{ExptRev10} employing non-invasive techniques. Their importance in fundamental Physics is crucial, providing a hotbed for studying the complex interplay of Coulomb-repulsion, quantum interference effects in the confinement, level quantization due to their smallness, and finally, the disorder in the form of irregularities in the confinement geometry.

The recent proposal of a insulator-metal transition in 2D electron gas (with inherent disorder)~\cite{RMP_Kravchenko} has renewed the interests for the study of ``disordered Wigner melting''~\cite{elecglass,chuitanatar95}, because such transition is often attributed~\cite{yoon} to a melting of a Wigner crystal or a Wigner glass. Disorder or impurities break all the spatial symmetries of the pure system reducing the ability of the particles to delocalize. There have been proposals for new and intervening phases~\cite{Waintal,Spivak,Sudip} between the crystal and liquid. Finally, the presence of disorder can change the nature and criticality of a transition~\cite{Aizenmann,Lebowitz}, and such modifications of Wigner melting have not yet been looked into. Overall, the complex interplay associated with such melting remains as one of the outstanding problems of condensed matter physics.

Seeking the physics of Wigner melting in confined geometries has significant experimental relevance~\cite{exp1} in varied areas, such as, radio-frequency ion traps~\cite{11Dyuti}, electrons on the surface of liquid He~\cite{12Dyuti}, electrons in quantum dots in semiconductor heterostructures~\cite{13Dyuti}, and in dusty plasmas~\cite{ExptRev10}. Unlike in the bulk system, the long-range Coulomb ($\sim r^{-1}$) repulsion is poorly screened in a finite cluster. Experimental clusters, such as large lateral quantum dots~\cite{MarcusGroup} often have inherent irregularities of the confinement, which are expected to act as disorder. This conjecture found support from the tunneling conductance measurements of chaotic dots~\cite{FolkMarcusGroup}, and associated Coulomb blockade experiments~\cite{Sivan96,cb99}. Although a quantum melting of Wigner-type in a circular trap has been studied extensively~\cite{Egger,Filinov01,Filinov08,Ghosal07}, a similar study in the irregular confinement has not yet received as much attention, see however Ref. \cite{ghosalhong05}. Recently, two of us reported the classical ``melting" of irregular Wigner molecule (IWM) under the influence of thermal fluctuations~\cite{Dyuti}, devised criteria to quantify such crossover, and identified its mechanism in terms of defects. There are fundamental questions to address upon inclusion of quantum fluctuations in the form of zero-point motion. How different is the nature of such melting in a IWM from that in confinements with certain symmetries? What roles do the defects play for such a quantum crossover? Finally, what would be the phase diagram in the plane of thermal and quantum fluctuations? Addressing such questions in the present paper, we list below our main findings:

\noindent (1) The symmetry of the confinement does not control the ``solid" to ``liquid" crossover scale in WM, neither does it influence the underlying ``melting" mechanism.

\noindent (2) The crossover turns out to be unique irrespective of the used distinct criteria for ``melting''.

\noindent (3) The phase diagram consists of three phases: {\em a classical liquid}, {\em Wigner molecule} and {\em a quantum liquid} along with intriguing {\em re-entrant solidification} with increasing temperature in the extreme quantum regime.

\noindent (4)
The crossover in the traps with spatial symmetries indicates that in the thermodynamic limit the quantum melting is likely to be of first order with the characteristic discontinuity in the observables. The weaker melting in an irregular confinement is more subtle for predicting thermodynamic limit. The thermal melting, on the other hand, follows the disclination mediated KTHNY mechanism~\cite{KT73,Halperin78,Young79,Nelson80}.

The rest of the paper is organized as follows. In Section II we introduce the model, the parameter space and the simulation method used to study both quantum and classical systems. In Section III, we present our main results. The corresponding subsections include the comparison of several melting criteria, the reconstructed diagram and the discussion of the melting mechanism. 

\section{Model, Parameters and Methods}
Our model of WM consists of $N$ distinguishable particles (Boltzmannons) in 2D plane and interacting via long-range Coulomb repulsion. These particles are trapped by an external potential $V_{\rm conf}({\bf r})$. For much of our calculations we compare results from three confinements: (a) Circular, $V_{\rm conf}^{\rm Cr}({\bf r})$, (b) Elliptic, $V_{\rm conf}^{\rm El}({\bf r})$ and (c) Irregular, $V_{\rm conf}^{\rm Ir}({\bf r})$, given by,
\begin{eqnarray}
V_{\rm conf}^{\rm Cr}({\bf r}) &=& \alpha r^2\nonumber \\
V_{\rm conf}^{\rm El}({\bf r}) &=& \alpha(x^2/c^2+ y^2/d^2)\nonumber \\
V_{\rm conf}^{\rm Ir}({\bf r}) &=& a\{ x^4/b+by^4-2\lambda {x^2}{y^2} +\gamma (x-y)xyr\}
\label{Vc}
\end{eqnarray}
where $r=\sqrt{x^{2}+y^{2}}$ and $\alpha = \frac{1}{2}m\omega_0^{2}$.
The confinement strengths $a$, $\alpha$ and $c,d$ controls the average particle density in these traps. Because we expect the $V_{\rm conf}^{\rm Ir}({\bf r})$  to mimic the universal features of disordered systems, it is designed to break all the spatial symmetries. For example, $b$ breaks the $x$-$y$ symmetry while $\lambda$ introduces chaoticity and $\gamma$ breaks reflection symmetry. Appropriate values for parameters representing universal disorder physics are found in literature~\cite{Bohigas93,Ullmo03,hong03,ghosalhong05}. While disorder is typically introduced in bulk systems by random impurities, they originate in a confined system with small number of particles from the irregularities at the `soft' boundary which are relevant for experiments in which the external confinement is set up by electrostatic and magnetic means~\cite{ExptRev10}.  The ensemble of particles in this potential will be referenced below as IWM. 

In case of a harmonic trap, the total Hamiltonian is given by 
\begin{equation}
H =  \sum_{i=1}^N \left[-\frac{\hbar^2}{2m}\nabla_i^2 + \frac{1}{2}m\omega_0^2 r_i^2\right] + \sum_{i<j}^N \frac{e^2}{\epsilon|{\bf r}_{i}-{\bf r}_{j}|},\label{ham}
\end{equation}
where $m$ and $\epsilon$ are the effective mass and the dielectric constant of the medium, respectively. Next, we introduce the length and energy scales $r_0$ and $E_0$, specified by $e^2/\epsilon r_0 = m\omega_0^2r_0^2 /2 $ and $E_0 = e^2/\epsilon r_0$. After the scaling transformation $\{ r \rightarrow r/r_0$  and $E \rightarrow E/E_0\}$ Eq.~(\ref{ham}) takes the dimensionless form~\cite{Filinov01} \\
\begin{equation}
H = \sum_{i=1}^N \left[-\frac{n^2}{2}{\nabla_i^2} + {r_i^2}\right] + \sum_{i<j}^N \frac{1}{r_{ij}}.
\label{h1}
\end{equation}
Here $n=\sqrt{2}l_0^2/r_0^2$ is the quantum parameter related with the strength of the trap defined via the oscillator length $l_0^2=\hbar/m\omega_0$. The analogy to the bulk in deciding the quantum parameter is given by $r_s = 1/\sqrt{\pi} n^2$, where $r_s$ specifies the average electron density, see Eq.~(\ref{rs}). The thermodynamic properties of~(\ref{h1}) have been studied in the canonical ensemble (fixed particle number $N$ and temperature $T=k_BT/E_0$).

The potentials $V_{\rm conf}^{\rm Cr}({\bf r})$  and $V_{\rm conf}^{\rm El}({\bf r})$ are quadratic in $r$, whereas, the $V_{\rm conf}^{\rm Ir}({\bf r})$  has quartic dependence. To match the dimensions of three potentials, we rescale the strength of the irregular potential as $a \rightarrow a'\left({m\omega^2}/{2r_0^2}\right)$. This rescaling ensures that $\{a',b,\lambda, \gamma\}$ in Eq.~(1)  can be treated as dimensionless parameters, whereas the definition of the quantum parameter $n$, being the pre-factor in the kinetic energy term, remains the same for all the three confinements.

The thermal melting of an irregular Wigner molecule has already been studied in detail~\cite{Dyuti}. In the current study we focus on the effect of quantum fluctuations (by varying $n$). Temperature $T$ is kept fixed at a low value, corresponding to the solid phase in the classical regime. By increasing $n$ we induce quantum melting in all the three confinements.

The quantum $N$-particle system~(\ref{h1}) can be analyzed by first principle simulations using the path integral Monte Carlo (PIMC) technique~\cite{Ceperley,BonitzBook}. The particle statistics is limited to ``boltzmannons'' (no exchange) to distinguish the effects related with the disorder and quantum statistics. In the path integral representation the $N$-particle density matrix is mapped on that of a classical system of interacting ``polymers''. Particles are represented by the trajectories which evolve in the imaginary time $0\leq \tau \leq \beta$, with the upper bound specified by the inverse temperature $\beta=1/k_B T$.
Number of time-slices $M$ changes with $\beta$ and the quantum parameter $n$ by the relation $M=l\beta n$, where $l$ is an integer in the range $1$ to $10$ to reach convergence in the energy better than $5\%$. In most of our simulations we use $M=100$ independent of temperature $T$. The ensemble averages have been estimated over $\sim 10^6$ independent realizations of particle trajectories efficiently sampled via the bisection algorithm~\cite{Ceperley}. 

Complementary, the classical counterpart has been studied by standard simulated annealing Monte Carlo (MC) scheme~\cite{cerny85} based on the Metropolis algorithm~\cite{metro53}. In the simulated annealing schedule we start from higher $T$ and anneal down to the desired $T$ to track the appropriate low energy states.

It is important to note, that the identification of the melting transition in the case of irregular traps~\cite{Dyuti} filled with quantum particles is more complicated compared with the classical clusters formed in circular symmetric potentials. The latter have been successfully studied via the modified Lindemann-based~\cite{Linde10} parameters, the orientational and radial correlation functions which take into account coexistence of two different types of symmetries, i.e. ordering into a triangular-lattice structure (with hexagonal symmetry) in the bulk of the cluster and a shell-structure in the boundary region~\cite{Lozovik1,BedPeet94,Lozovik2}, as wells as possible radial inhomogeneity of the onset of melting.

In contrast, the onset of melting in quantum clusters will be significantly masked by the zero-point fluctuations. In particular, the Lindemann-based parameters and the correlation functions in quantum solids at $T=0$ get a significant background shift related with the quantum mechanical uncertainty in the particle position. As an example, the critical value of the Lindemann parameter at the solid-liquid quantum transition~\cite{Ast} ($T=0$) is increased to $0.2$ from its original value $0.1$ valid for classical solids. Similar observations hold for finite quantum clusters~\cite{Filinov01,Filinov08,Bonitz}. The combined effect of quantum fluctuations and the order-disorder transition is even more difficult to distinguish in small systems ($N\lesssim 100$) where a sharp transition is replaced by a crossover behavior.

Partially to overcome this problem in the present studies, we measure the order parameters characterizing the ``solid phase'' in two different ways. First, by taking the ensemble average over particle configurations on every imaginary time slice. This is called {\em bead-by-bead} (BB) distribution and corresponds to the correct quantum mechanical average. Second, we exclude the zero-point fluctuations related with the delocalization of quantum particles and address the {\em bead-centroid} (BC) positions during the imaginary time evolution
\begin{align}
 \langle{\bf r}_i\rangle=\frac{1}{\beta}\int_0^{\beta} {\bf r}_i(\tau)d \tau, \quad i=1,N.
 \label{bc}
\end{align}

The physical picture is captured the BB-distribution as it takes into account the wave function overlap. The BC-distribution, instead, is more suited to describe different structural symmetries, cf. relative particle positions on the shells, and corresponds to the semi-classical picture. The BC-picture also depicts the evolution of particles in the configuration space, while the BB the combined evolution including the imaginary time. 
Because BC-calculations provide a semi-classical interpretation for the quantum effects, it is useful for the comparison between the classical and its corresponding quantum counterpart.
Both pictures are necessary to fully understand different melting criteria introduced below.

\section{Results}

In our studies we mainly concentrate on the system with $N=57$ particles. In the case of circular potential it is called {\em Magic Cluster} as the hexagonal symmetry dominates over the full structure~\cite{BedPeet94}. In irregular and elliptical case, no such concept as a magic cluster exists. 

Next, we should address the question: which parameter needs to be fixed to compare melting in three different traps on equal grounds? The parameter $n$ is inversely proportional to $r_0$, the spacing between two nearest neighbors. We tune the trap parameters ($c$ and $d$ in the elliptical trap, and $a'$ in the irregular one) to make $r_0$ equal in the three cases. This results in the same average density. To do this, we compare the coordinate $r_0$ for the ground state of two particles obtained by minimizing the potential for the different traps. 

The trap parameters found by this procedure are: $a'=0.0954$ in case of {\em irregular}, and $c=1.25$ and $d=0.85$ in case of {\em elliptical} potential. The other parameters for the {\em irregular} trap are similar to those used in Ref.~\cite{Bohigas93}, cf. $b=\pi/4$, $\lambda=0.635$ and $\gamma=0.2$. How close the spacing $r_0$ approximates the average inter-particle distance can be checked from the pair distribution functions presented below. 

\begin{figure}[t]
\includegraphics[width=9.50cm,height=10.50cm]{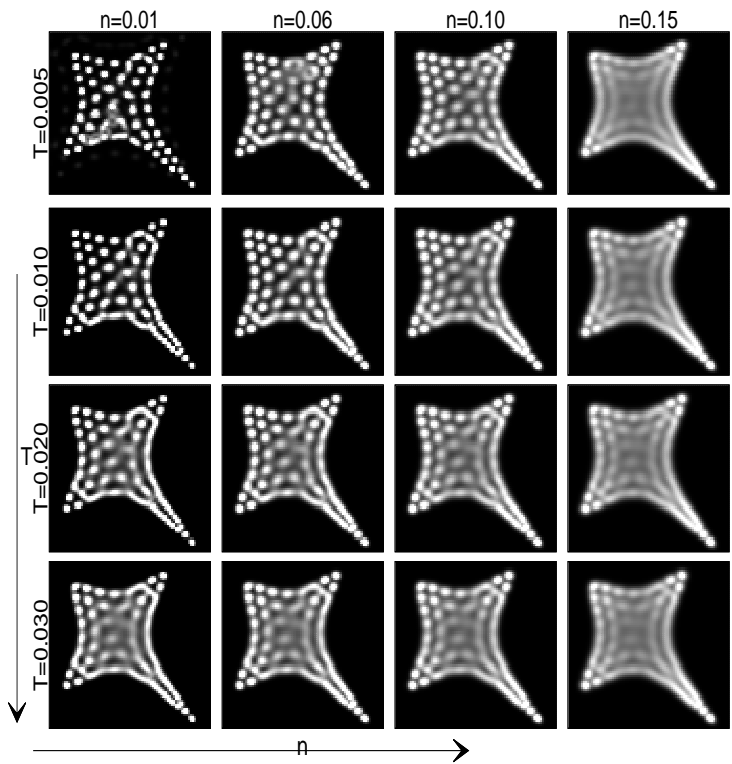} 
\caption{
2D density $\rho(x,y)$  plots on the $T$-$n$ plane (temperature-quantum parameter). Quantum fluctuations $n$ are increased from left ($n=0.01$) to right ($n=0.15$) column. Following along a column (fixed $n$) or a row (fixed $T$) one can trace the melting induced either by thermal or quantum fluctuations. Note, that due to the combined effect of the external field and pair interactions the density fluctuations are strongly anisotropic.
A typical cluster size $L/r_0$ can be read out from Fig.~\ref{fig9}. Our length unit scales as $r_0 \propto 1/n^2$. Thus, while the rescaled system size stays practically unchanged with $n$, its absolute spatial dimensions evolve quite rapidly with $n$.
}
\label{fig1}
\end{figure}

\subsection{Study of the crossover from density profiles}

Clear evidence on the melting transition in finite systems (``solid''-``liquid'' crossover) can be directly gained from two-dimensional density distributions. Fig.~\ref{fig1} presents a sequence of density plots for the IWM in a given confinement.  

The left column presents an example of melting by thermal fluctuations (the quantum parameter has the smallest value $n=0.01$).
Here, fluctuations are nearly frozen at low $T$ and $n$ and individual particles are resolved as bright spots in the density profile (though certain particles show quantum signature through a spread in their wave functions). With the increase in $T$, particle delocalization occurs in two ways: (a) Thermal diffusion of particles around their equilibrium position, and (b) The thermal fluctuations create occasional long and string-like paths of delocalization, whose role on the crossover is already discussed in the literature~\cite{Dyuti}.
Moving along a row, (say, the one corresponding to $T=0.005$), the strength of the quantum fluctuation increases and the corresponding crossover occurs primarily by the spreading of wave functions of the individual particles. We found no additional `string-like paths' upon increasing $n$. However, such paths, once developed for $T \geq 0.01$, survive quantum fluctuations up to a large $n$ and the melting becomes complete when such paths cannot be resolved from the spread of particles' wave function. This is clearly seen in the third row from the top, corresponding to $T=0.020$.

Now let us draw attention to the $N$-particle wave function (or density) at $n=0.15$ and $T=0.005$. Here the individual particle positions are hardly distinguished resulting in a quantum liquid state.
Interestingly, a rise in the temperature to $T=0.010$ leads to slight but noticeable particle localization in the center and the first row. This illustrates a gradual transition from quantum to quasi-classical picture. Maximum of both thermal and quantum fluctuations is reached in the bottom right panel (in our parameter space) and represents a `melted' IWM.

While the qualitative picture of the crossover is apparent from Fig.~\ref{fig1}, few points deserve special mention: (a) In the quantum cluster at $n=0.15$ the particles are reasonably delocalized even at the lowest $T$. However, heating to $T=0.020$ seems to weaken the ``liquidity'' and distinction of individual particles becomes better than for $T=0.005$. This is in contrast with the scenario common for low $n$ ($n=0.01$), where the temperature increase `dissolves' the particles monotonically. This observation indicates emergence of the incipient {\em re-entrant} behavior showing weakening ``liquidity'' with increasing $T$. A final classical liquid is obviously expected for $T \rightarrow \infty$. This feature is qualitatively consistent with the similar trend observed in bulk systems, where re-entrant crystallization from a quantum liquid phase by increasing the temperature was reported~\cite{casula}; (b) As can be clearly seen from Fig.~1 all particles do not melt simultaneously upon increasing the fluctuations.
The particles on the `string-like path' diffuse in the background more easily, because, the spreading of wave function is easier as these particles undergo larger thermal displacements;
 
(c) In our rescaled units the average inter-particle distance stays practically unchanged with variation of $n$, i.e. $\Avr{r}/r_0\sim 1$. This also holds for the cluster size, see Fig.~\ref{fig1}. However, in the absolute units $\Avr{r}$  shows a strong $n$-dependence: $\Avr{r}\sim r_0 =a_B/n^2$. Similar dependence is captured by the quantum Bruckner parameter
\begin{align}
r_s = [\pi \rho a^2_B]^{-1/2}=\Avr{r}/[a_B \sqrt{\pi}]=1/[\sqrt{\pi}n^2],
\label{rs}
\end{align}
typically used  to characterize a bulk density of homogeneous systems. Finally, Eq.~(\ref{rs}) can be used for comparison of the onset of melting in bulk and trapped systems.

\begin{figure}[t] 
\includegraphics[width=9.5cm,height=10.50cm]{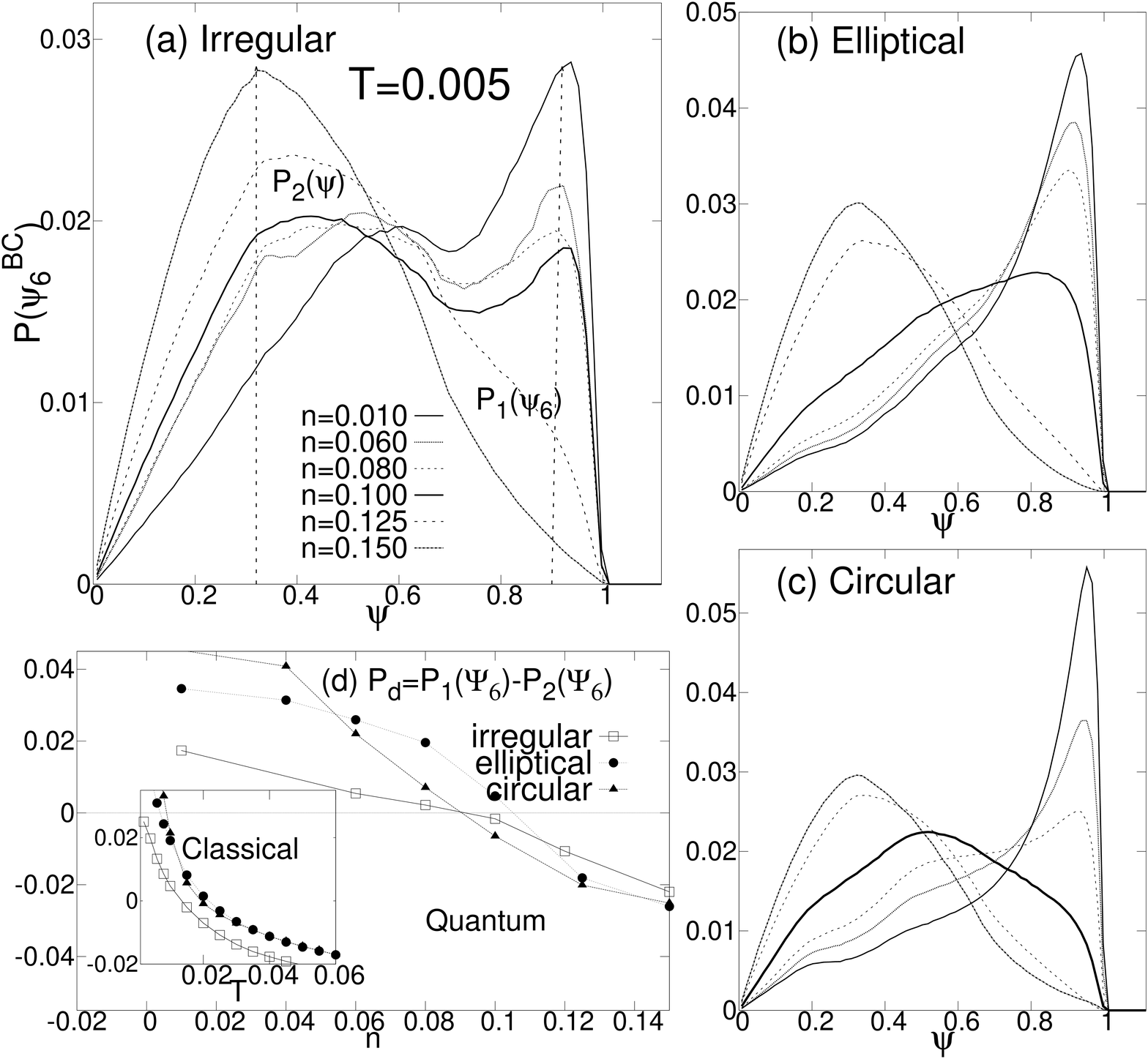}
\caption{
(a-c) Distribution of the bond-angular order parameter $P(\psi_6)$ shown for different strength of quantum fluctuations $n$. $\Psi_6$ is evaluated from  the bead-centroid (BC) particle coordinates $\langle {\bf r}_i \rangle$, Eq.~(\ref{bc}). Three different traps (a)-(c) are analyzed at fixed temperature $T=0.005$. Common behavior is observed: the maximum, $P_n=\max[P(\psi_6)]$ ($n=1,2$), gradually shifts from the position $\Psi_6\sim 0.9$ specifying a ``solid'' phase to $\Psi_6\sim 0.3$ typical for a quantum or classical ``liquid''.
The weakness of bond-orientation order in the irregular trap reflects in the weak peak at $P_1$ even at the lowest $n$, and $P(\psi_6)$ in an irregular trap features a second peak drifting towards $P_2$ with increasing $n$.
d) The $n$-dependence of $P_d(n)=P_1-P_2$ during the ``solid''-``liquid'' transition. $P_d(n)$ demonstrates a similar trend for all three traps and changes its sign around $n_c\approx 0.10(1)$. The inset shows the evolution of $P_d$ in a similar classical systems. Here it is concave in nature while in the quantum case it is of convex nature.
}
\label{fig2}
\end{figure}

\subsection{Crossover in bond orientational order}\label{boo}
The crossover in finite systems bringing out the correlation driven Wigner physics, naturally warrants proper qualification.
Here we study the evolution of the bond orientational order (BOO)~\cite{Nelson79,Nelson83} with $n$. The hexagonal symmetry and deviation from it can be quantified by the order parameter
\begin{equation}
\Psi_6=\left\langle \psi_6({\bf r}) \right\rangle=\left\langle\frac{1}{6}\sum_{nn=1}^{6}\exp(6i\theta_{nn}({\bf r}))\right\rangle,
\label{psi6}
\end{equation}
where the sum is taken over six nearest neighbors ($nn$) of a particle located at the position ${\bf r}$, and $\theta_{nn}$ is the relative angle between
the vector ${\bf r}_{nn}-{\bf r}$ and an arbitrary fixed axis. The ensemble average is taken over all particle positions.

Evaluation of $\Psi_{6}$ has been performed using the BB- and BC-methods (see above). In the first case, the BOO includes quantum mechanical fluctuations, and as a result $\Psi_{6}$ is significantly reduced in a quantum solid even with a perfect hexagonal symmetry. In the bead-centroid (BC) method the averaged particle positions~(\ref{bc}) are used, allowing a semi-classical interpretation and demonstrating the behavior typical for classical solids.

Fig.~\ref{fig2} compares the distribution of the order parameter $\Psi_6\equiv\Psi_6^{\text{BC}}$ for three different confinement potentials, several values of $n$ and lowest temperature $T=0.005$. In the ensemble average~(\ref{psi6}) we excluded the particles at the cluster boundary. 

A comparative study of $P(\Psi_{6})$ in the three confinements brings out their common qualitative evolution:
At small value of quantum parameter ($n=0.01$) $P(\Psi_{6})$ is peaked around $\Psi_{6}=0.9$ indicating that in the cluster bulk the majority  of particles have six $nn$ with the hexagonal symmetry. As $n$ is increased the peak weakens and shifts progressively to lower values. Finally, at $n=0.15$ we observe a rather broad distribution centered around $\Psi_{6}=0.3$. Further increase of $n$ does not significantly alter this shape.
Inspite of these broad similarities, the evolution of $P(\Psi_{6})$ in the irregular trap differs from those in a circular or elliptic confinement: Each trace of $P(\Psi_{6})$ features double peaks up to $n \sim 0.10$ and the one at $\Psi_{6}=0.9$ is weaker than those in the traps with spatial symmetries. The second peak, far weaker than the first one, drifts towards $\Psi_{6}=0.3$ contributing to the sole peak at large $n$.

The convergence of the distribution $P(\Psi_{6})$ to one characteristic shape for all confining geometries at large $n$
can be explained by the fact that any peculiarities of an external potential are effectively screened due to quantum delocalization of particles. This induces the effect of smoothness in the external field as has been demonstrated by the variational perturbation theory due to Feynman and Kleinert~\cite{kleinert}. The part of a full partition function related with an external field can be substituted by an effective classical potential which accounts for possible effects of quantum fluctuations. The induced smoothness, also in the pairwise interactions, is well known in the literature~\cite{fil2004}. 

In the limit of small $n$ we also observe quite universal form of $P(\Psi_{6})$ with the peak at $\Psi_{6}=0.9$. This again comes from the screening effect but classical in its origin. It is specific to Coulomb systems, that particles mainly experience interaction with their nearest neighbors, whereas the interactions at larger distances coming from the outer parts of the system are canceled. The cancellation will be complete in the mean field approximation.

Another common feature found for all three traps is the bimodal nature of the distribution. The height of $P(\Psi_{6})$ at two reference values,  $\Psi_{6}=0.9$ and $\Psi_{6}=0.3$, which distinguish ``solid'' and ``liquid'' phases, can be used to characterize the degree of ``solidity'' and ``liquidity'' in the crossover regime. We have evaluated the corresponding heights $P_1=P(0.9)$ and $P_2=P(0.3)$ and their difference $P_d$ for $0.01 \leq n \leq 0.15$. 
We plot $P_d$ in Fig.~\ref{fig2} for all three types of the confinement and observe a qualitatively similar behavior. The $n$-dependence of $P_d$ demonstrates a monotonic decay from positive to negative values. This allows to identify the crossover from a ``solid" to a ``liquid" phase. The point $n_X \sim 0.10(1)$ [which corresponds to the bulk density $r_s \sim 56(10)$ using Eq.~(\ref{rs})] where $P_d$ changes the sign is the mid-point of the crossover. Similar to our previous discussion of the BOO, the nature of the confinement does not play a significant role in the transition induced by quantum fluctuations.

The inset in the Fig.~\ref{fig2} presents similar analyses for the classical system in the same confinement (of all three types) where the transition is triggered by thermal fluctuations. The comparison shows two distinct differences. First, although the evolution of $P_d$ with $n$ (at fixed $T$) is of convex nature, $P_d$ versus $T$ (at $n=0$) has a concave curvature. Second, the BOO in $V_{\rm conf}^{\rm Ir}({\bf r})$ is destroyed earlier compared to $V_{\rm conf}^{\rm Cr}({\bf r})$  and $V_{\rm conf}^{\rm El}({\bf r})$ . Both these features can be understood by connecting the mechanism of melting with the defect production, which will be discussed in the details in Section~G. 

For completeness we also calculated $P(\Psi_6)$ from~(\ref{psi6}) using the BB-distribution, $\Psi_6\equiv \Psi_6^{\text{BB}}$. In this approach we capture additional (background) fluctuations from the individual particle trajectories in the imaginary time. Although $P(\Psi^{\text{BB}}_{6})$ shows similar trend as $P(\Psi^{\text{BC}}_{6})$ its larger background fluctuations makes the identification of the crossover more difficult.

\begin{figure}[t]
\includegraphics[width=9.5cm,height=10.50cm]{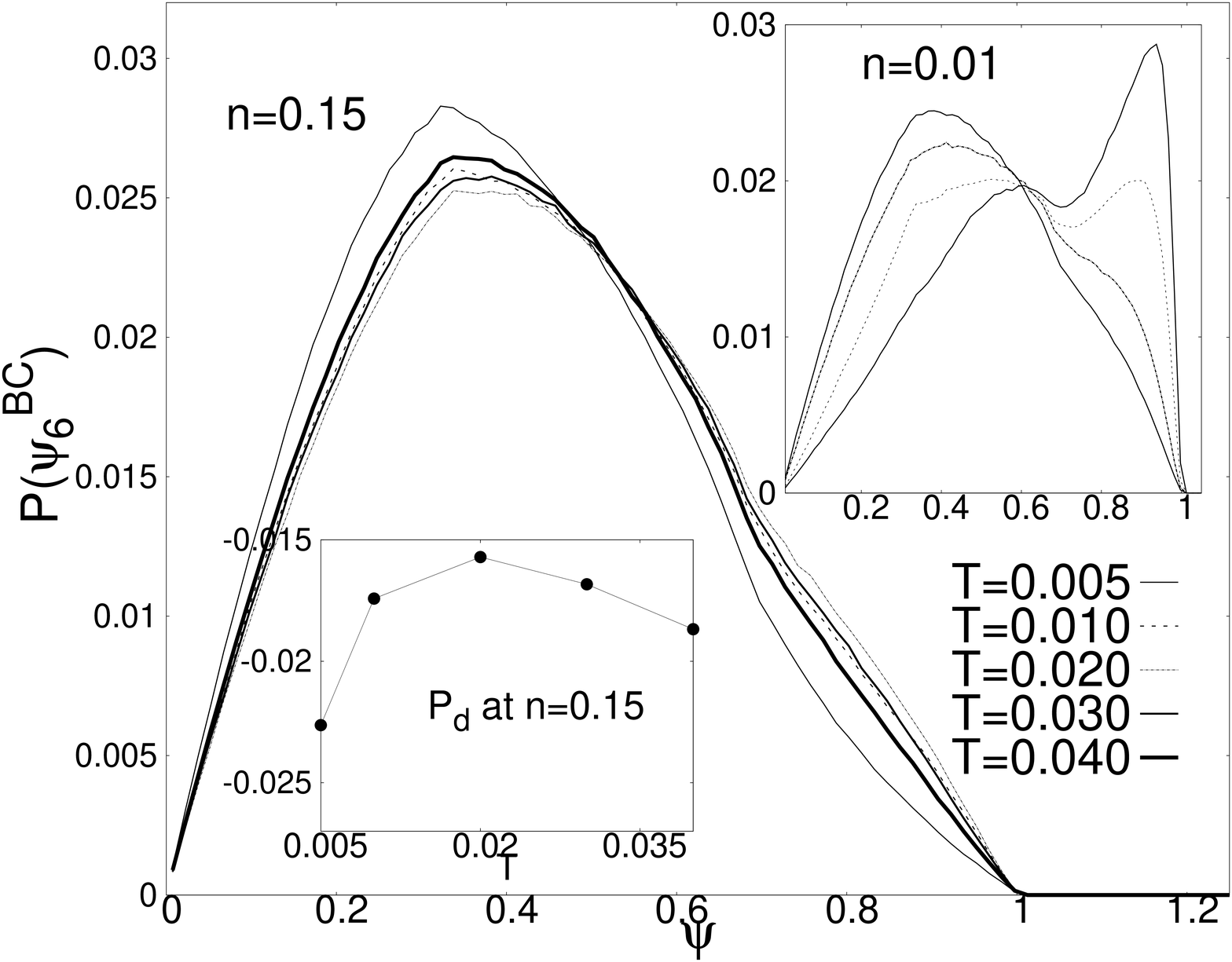}
\caption{
{\em Main panel}: $T$-dependence of $P(\Psi_6)$ deep in the quantum regime ($n=0.15$). The re-entrant behavior is clearly observed 
for $0.01\leq T\leq 0.02$. {\em Lower inset}: $T$-dependence  of the height difference $P_d$. The non-monotonic behavior is observed in the same temperature range and is related with a partial restoration of the hexagonal order. {\em Upper inset}: Typical behavior of $P(\psi_6)$ in the quasi-classical case ($n=0.01$). A similar dependence is observed in the classical limit ($n=0$).
}
\label{fig3}
\end{figure}

It is interesting to check whether the {\em re-entrant behavior}, as observed from the density plots in Fig.~\ref{fig1}, can be more quantitatively analyzed via the BOO. In Fig.~\ref{fig3} we present the $T$-dependence of $P(\Psi_6)$ at the largest value of quantum parameter $n$. In this case, surprisingly, the ``liquidity'' of the system defined in terms of the height of the distribution $P_2=P(\Psi_6)|_{\Psi_6=0.3}$ dominates at low temperatures. As the temperature is increased from $T=0.005$ to $T=0.04$ the height $P_2$ decreases, whereas the right wing of the distribution around $\Psi_6=0.9$, characterizing the degree of ``solidity'' (or hexagonal symmetry), steadily increases. This leads to the conclusion that the hexagonal order is higher at $T=0.04$ then $T=0.005$. This finding can be explained by the anisotropy of quantum fluctuations at low temperatures. Quantum particles are mainly delocalized along narrow potential valleys, following the shape of the trap, as shows the comparison of the density plots in Fig.~\ref{fig1} for $n = 0.10$ and $n = 0.15$ at $T = 0.005$. This naturally leads to a significant distortion of the BOO as we approach the cluster boundary, where the symmetry of the irregular potential dominates over the hexagonal order. In contrast, at high temperatures the particles get more localized and the BOO is dominated by the bead-centroid particle coordinates~(4) which resemble the hexagonal order with some distortions. As a net result of this transition to the quasi-classical regime the hexagonal symmetry is partially restored.

The lower inset in Fig.~\ref{fig3} shows the difference of the peak heights $P_d$ of two ``phases''. Starting from negative values, due to a loss of the hexagonal order in the boundary region at low $T$, $P_d$ shows a non-monotonic $T$-dependence. It increases for the temperature range $0.01 \leq T\leq 0.02$ being a clear indication of the weakening of ``liquidity'' and resembling the {\em re-entrant behavior} observed in the macroscopic systems~\cite{casula}. However, we should stress that in our case this feature is observed due to a strong perturbation of the BOO in the boundary region at low temperatures.
As discussed above the BOO is restored at high temperatures. The prominence of this effect is naturally expected in finite size systems where the boundary effects play an important role.

Finally, we analyze the evolution of the BOO in the quasi-classical regime where the thermal fluctuations make important contributions.
The upper inset in Fig.~\ref{fig3} shows the $T$-dependence of $P(\psi_6)$ at $n=0.01$. We find the standard behavior as observed in Fig.~\ref{fig2}: during the crossover $P_d$ decreases monotonically and there is no sign of the {\em re-entrant behavior} in $P(\psi_6)$.
    
\subsection{Study of the crossover via the Lindemann ratio}\label{lind}
An additional quantity which tracks the solid-liquid crossover is the Lindemann parameter~\cite{Plasma12,Linde10} 
\begin{equation}
{\cal L}^{\rm \text{BB}} = \frac{1}{N M} \sum_{i=1}^{N} \sum_{t=1}^{M} \sqrt{\frac{\langle r_{it}^{2}\rangle}{\langle r_{it}\rangle^2}-1}.
\end{equation}
In the ensemble average $\langle \ldots \rangle$ we use the bead-by-bead distribution of particle trajectories from the PIMC simulations. Here $M$ is the total number of time slices and ${\bf r}_{it}$ is the position of the $i$-th particle at the $t$-th time slice relative to the trap center. 

\begin{figure}[t]
\includegraphics[width=9.5cm,height=10.50cm]{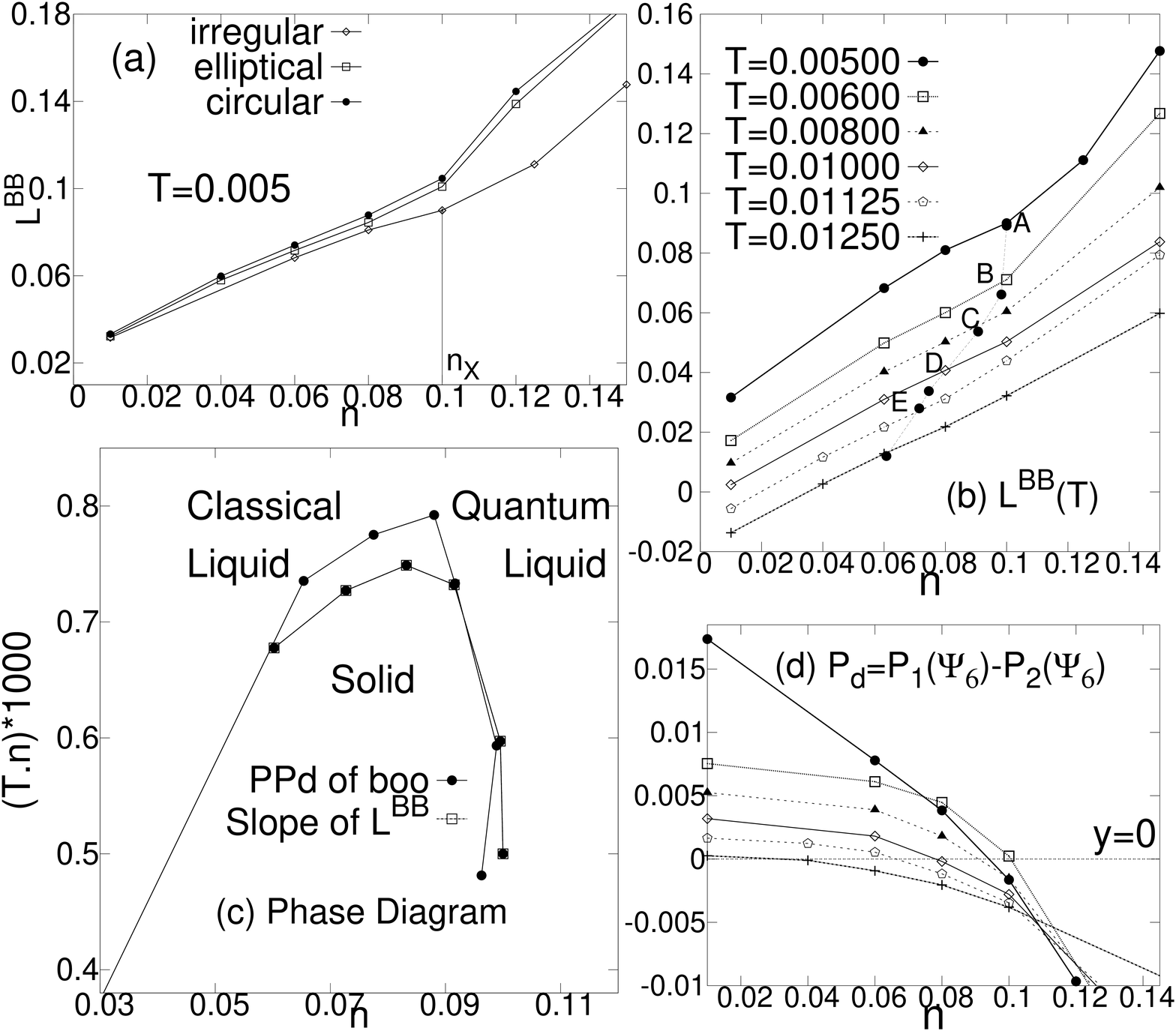}
\caption{(a) Evolution of ${\cal L}^{\text{BB}}$ with $n$ for three different traps. In all cases we observe a similar trend with a characteristic kink which distinguishes two slopes in the evolution of ${\cal L}^{\text{BB}}$ with $n$, and defines $n_X(T)$. Such differences in the slope become weaker with $T$.
(b) The locus of $n_X(T)$ traces the curve ABCDEF and translates into the phase diagram in the $T-n$ plane shown in (c). Similarly the evolution of $P_d$ with $T$ at (d) gives more confidence to draw the phase diagram. (c) Phase diagram: two curves representing $n_X(T)$ (estimated from two independent criteria) are compared and distinguish different phases.}
\label{fig4}
\end{figure}

In Fig.~\ref{fig4}(a) we show the evolution of ${\cal L}^{\text{BB}}$ with $n$ at lowest temperature ($T=0.005$) for all three traps. First, ${\cal L}^{\text{BB}}$ evolves linearly with $n$, and a distinct change in the slope takes place only around $n_X \approx 0.11(1)$. This holds for all three cases validating that the quantum crossover parameter $n_X$ is universal for different trap geometries, though the change of slope in the irregular confinement is rounded off beyond $n_X$. We note that ${\cal L}^{\text{BB}}$ demonstrates a strong dependence on the quantum fluctuations $n$ even in a solid ``phase''. This is in contrast to the findings for classical systems, where only a weak dependence on thermal fluctuations was observed below a critical temperature ($T<T_X$)~\cite{Dyuti,BedPeet94}. Our observation is not surprising, as in the quantum case the Lindemann parameter captures the background contribution coming from the evolution in the imaginary time and the quantum mechanical uncertainty. Both are growing with $n$. In fact, ${\cal L}^{\text{BC}}$ (not shown here) that picks up only a semi-classical contribution, shows more resemblance with the classical behavior as expected.

\subsection{Phase Diagram}
Our detailed analyses of ${\cal L}^{\text{BB}}$ versus $n$ for different temperatures, cf. Fig.~\ref{fig4}(b), allows us to generate a phase diagram, cf. Fig.~4(c). The locus of $n_X(T)$ traces the curve ABCDEF in Fig.~\ref{fig4}(b) and is translated into the phase boundary in the $T-n$ plane. 

To strengthen our confidence on the general validity of the phase diagrams reconstructed based on the Lindemann parameter~\cite{Filinov01}, we perform, in addition, an independent reconstruction using our $P_d$-criterion, cf. Fig.~\ref{fig2}. The critical points $n_X(T)$ are defined by the sign change in $P_d$. As an example, Fig.~\ref{fig4}(d) shows $P_d$ versus $n$ for several temperatures $T$ in the case of the $V_{\rm conf}^{\rm Ir}({\bf r})$. 

The phase boundaries reconstructed from both criteria demonstrate good agreement, cf. Fig.~\ref{fig4}(c). We can conclude, that the use of both methods for the melting-analyses is justified and mutually complementary.

Next, we discuss some important features of the phase diagram. The left-most side of the phase boundary (for $n \lesssim 0.05$) corresponds to a `quasi-classical' regime where the degree of ``liquidity'' is mainly due to thermal fluctuations. In contrast, quantum fluctuations play a dominant role for the right-most phase boundary and induce the zero-temperature melting around $n_X\sim 0.11(1)$.

The slope of the phase boundary for small $n$ can be well described by a single coupling parameter $\Gamma$ well known for classical systems. The tangent
is given by $\Gamma^{-1}=k_B T_{cl}/\langle V \rangle$, i.e. the classical melting temperature measured in units of the characteristic potential energy $\langle V \rangle$. Quantum fluctuations systematically reduce $T_{cl}$, as the system deviates progressively from the classical limit, making the boundary line sub-linear in $n$. Deep in the quantum regime, when the zero-point fluctuations dominate, the curve bends down fast. The maximum in the melting temperature is reached for some intermediate $n$-values. Remember, that $n$ was defined as the reduced density parameter scaling as $n\propto 1/r_0^2$. Hence, the absolute value of the critical temperature $T_c$ shows a non-monotonic behavior. First, $T_c$ is increased with the density. Then, above some critical density ($n^{\star}\sim 0.09$), it fast reduces to zero. Clearly, the observed peak in the reduced temperature $T=k_B T/\langle V \rangle$ around $n^{\star}$ occurs at $T < T_{cl}(n^{\star})$.

\subsection{Correlation functions}
We further quantify the solid-liquid crossover and the different phases from the study of the spatial pair correlation function~\cite{Ott11}
\begin{equation}
g(r) =\frac{1}{N} \left\langle\sum\limits_{i=1}^N \sum\limits_{j>i}^N \delta(r-|{\bf r}_i-{\bf r}_j|)\right\rangle, 
\label{gr}
\end{equation}
and the bond-orientational correlation function
\begin{equation}
g_{6}(r)=\left\langle\psi_{6}^{\star}(|{\bf r}^{\,\prime}|)\psi_{6}(|{\bf r}^{\,\prime}-{\bf r}|)\right\rangle.
\label{g6}
\end{equation}
The spatial decay of $g_6(r)$ characterizes the correlation length of local orientational order ${\psi_6({\bf r})}$, while $g(r)$ describes the probability to find any pair of particles at distance $r$.

\begin{figure}[t]
\includegraphics[width=9.0cm,keepaspectratio]{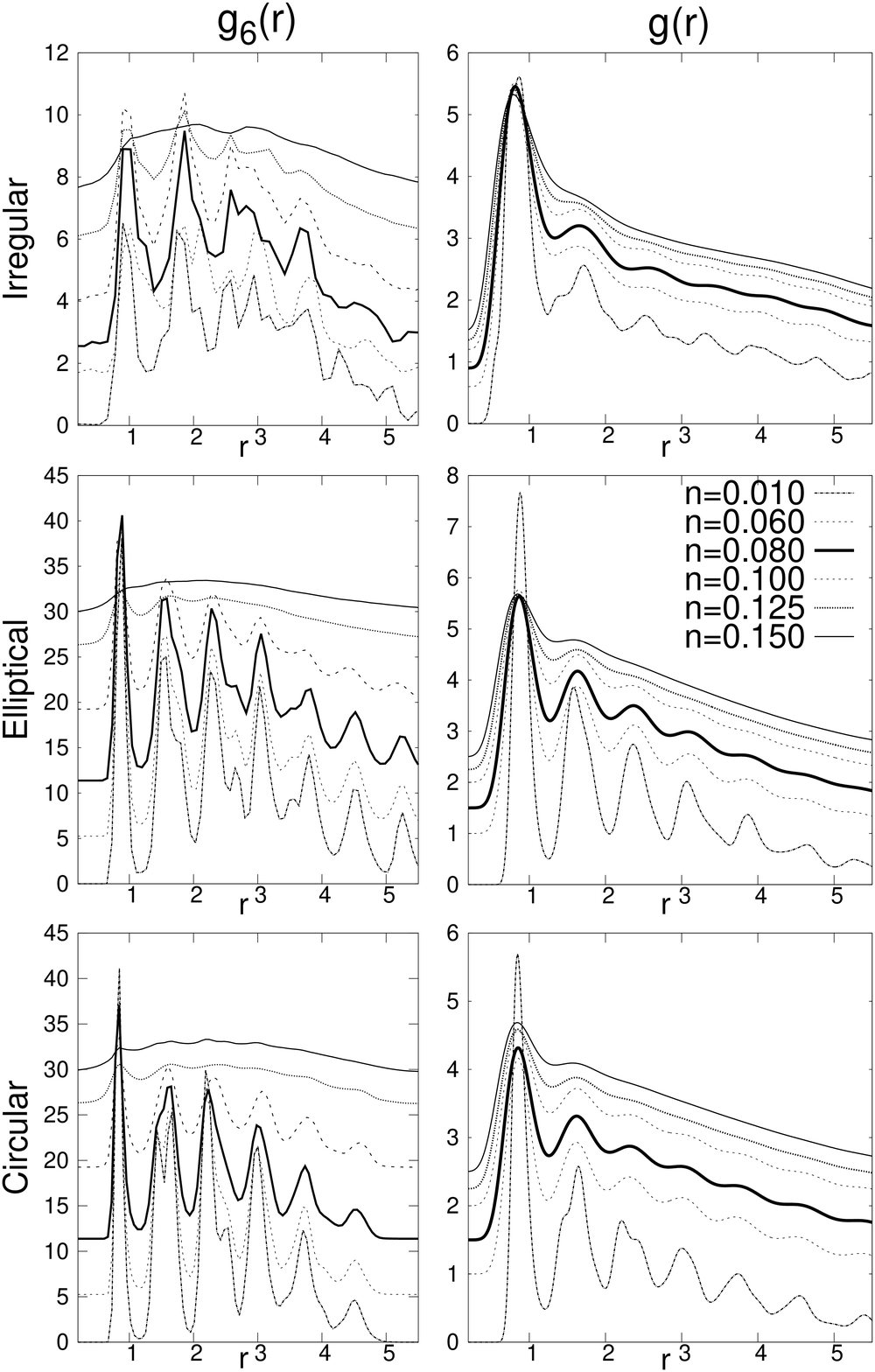}
\caption{
{\em Left column}: Evolution of $g_6(r)$ versus $n$ estimated with the BC-method for three different confinements. {\em Right column}: Evolution of the pair correlation function $g(r)$. The legend indicates the $n$-values used in all plots. The center of the crossover is identified with $n_X=0.10$ when the Bragg peaks in $g_6(r)$ disappear. Similar observation holds for all three traps.
}
\label{fig5}
\end{figure}

The amorphous solids (as IWM) is characterized by its quasi-long range orientational order and a depleted positional order~\cite{Dyuti} in the limit of $n,T \rightarrow 0$. It is interesting to study how these correlation functions evolve with $n$, i.e. in the presence of quantum fluctuations. Such results are presented in Fig.~\ref{fig5}. The function $g(r)$ was calculated by the BB-method,  while $g_6(r)$ using the BC-method [in Eq.~(\ref{g6}) we substitute $\psi_6({\bf r})\equiv {\psi^{\text{BC}}_6({\bf r})}$]. 

Our results for $g_6(r)$ (left column in Fig.~\ref{fig5}) show well defined Bragg peaks up to a distances comparable with the linear dimension of the system, indicating the long-range nature of the BOO in all three traps. By increasing $n$, the Bragg peaks broaden progressively up to $n_X \approx 0.11$. Then they disappear abruptly leaving bump-like features of a liquid. This is consistent with the crossover scale inferred from other criteria introduced in Sec.~\ref{boo} and~\ref{lind}. The long range nature of these correlations and their sudden demise are independent of the nature and symmetry of the confinement. 

For the circular and elliptic confinements $g(r)$ provides sufficient information on the positional order. The Bragg peaks of such a mesoscopic ``solid'' phase can be well resolved. The disappearance of the peaks with quantum fluctuations is smooth and is in contrast to the behavior of $g_6(r)$. 

Slightly different behavior is observed for the irregular trap. Here, the peaks in $g(r)$ are broader and overlap even for small $n$-values indicating that there is no positional order. 

Finally, we note that the location of the first peak in $g_6(r)$ and $g(r)$ is close to one (in units of $r_0$) irrespective of the confinement type. This validates our choice of parameters for the trapping potentials to ensure the same average density.

\subsection{Signature of the classical/quantum crossover in the fraction of defects}

Thermal melting in 2D follows the KTHNY~\cite{KT73,Halperin78,Young79,Nelson80} theory and demonstrates some universal features mediated by the unbinding of topological defects. It is the dislocations and disclinations~\cite{kleman} that signal the destruction of the positional and orientational orders, respectively. The irregularities in the confined systems, if any, are known to mask the proliferation of the free dislocations~\cite{Dyuti}, but the disclinations still play a major role in the thermal melting. The natural question thus arises: What role do the disclinations play, if at all, for the quantum melting in confined geometries?

We address this question by analyzing the fraction of particles $N_{C_n}/N$ with the coordinate numbers ${C_n}$ ($n=4,\ldots 8$) for a wide range of the quantum parameter $n$ and compare the results with thermally induced melting. 

\begin{figure}[t]
\includegraphics[width=9.5cm,height=9.50cm]{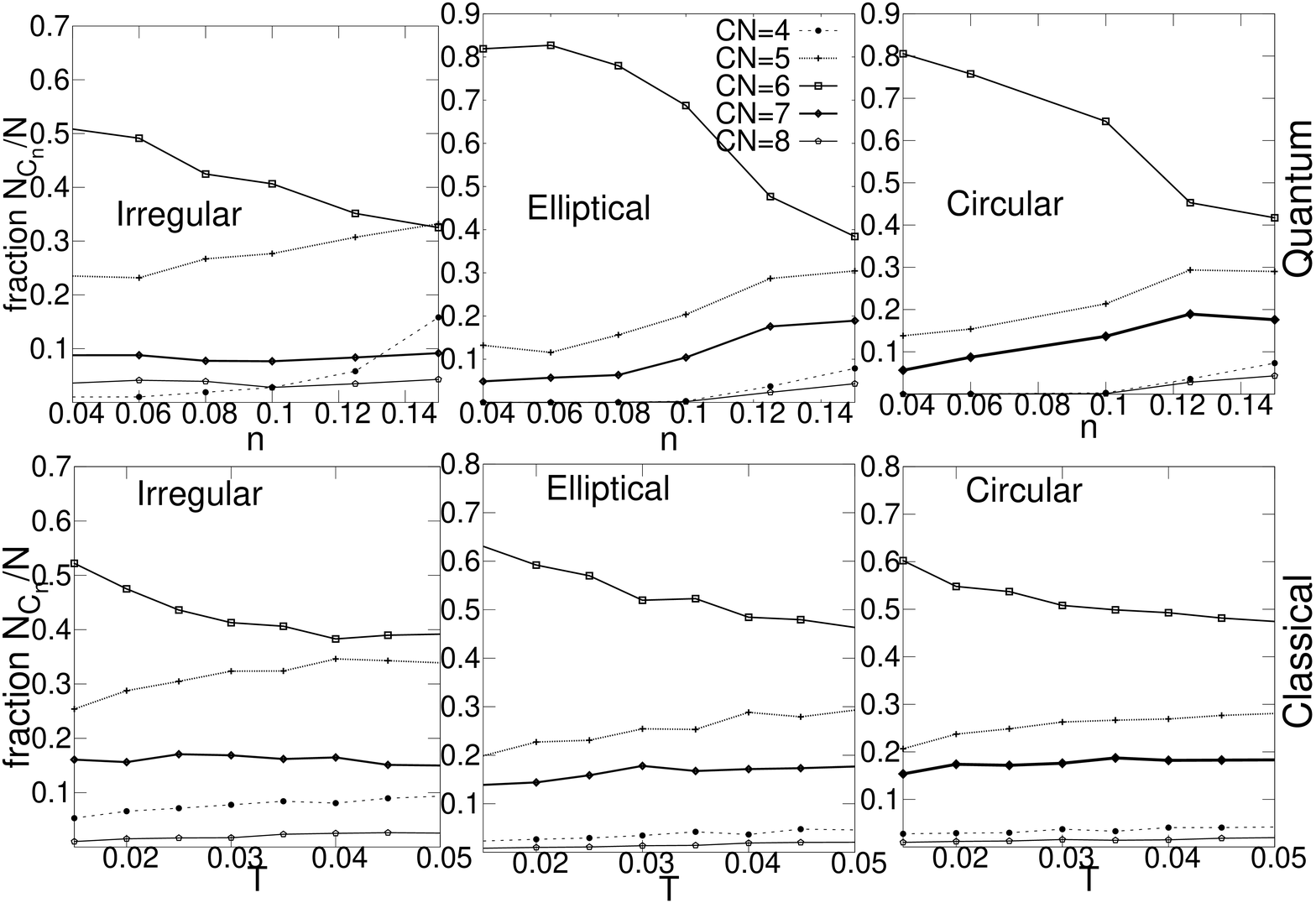}
\caption{
Fraction of the defects versus the tunning parameter: $n$ in quantum (upper row) and $T$ in classical (lower row) cases. Qualitatively similar trends are observed: the particle fraction, $N_{C_n}/N$, with the coordinate numbers $C_n$ $(n=5,7)$ steadily increases with both thermal and quantum fluctuations. The decay of the hexagonal order, cf. $N_{C_6}/N$, is of convex nature in the quantum case though the curvature is rather weak for the irregular confinement as well as its magnitude at the lowest $n$. In contrast, its evolution is of concave nature in the classical case. The mechanism behind melting in the classical system is the proliferation of disclinations. It is not conclusive for the quantum case due to the evolution in the imaginary time and effective presence of third dimension.
}
\label{fig6}
\end{figure}

The particle numbers $N_{C_n}$  have been unbiased estimated by the Voronoi construction~\cite{voronoi88}. The results for the three traps are presented in Fig.~\ref{fig6}: quantum melting (upper row) and thermal melting (lower row). In both cases, $N_{C_6}$ makes the largest contribution in the ``solid'' phase for all the traps. This fraction, however, is steadily decreased by both thermal and quantum fluctuations in the expense of generating new disclinations, i.e. increase of the populations $N_{C_n}$ with $C_n \neq C_6$. 

The fractional change in the number of these defects
for circular or elliptic confinements
are more abrupt as a function of $n$ (upper panel) compared to their gradual evolution with $T$ (lower panel).
The convex (or concave) nature of the evolution of $P_d$,
at least for the circular and elliptic traps
in Fig.~\ref{fig2} in quantum (or classical) case can be understood by the nature of evolution of the defect-fraction.
A similar argument for the irregular trap is masked by a rather weak $P_d$, even deep in the ``solid", and makes such a correlation between the shape of $P_d$ and $N_{C_6}$ less substantial.
The lack of symmetry in the $V_{\rm conf}^{\rm Ir}({\bf r})$ leads to a relatively fast destabilization of six-coordinated neighbor which makes $P_d$ to approach zero at a slightly lower $T$ and also a lower $n$ than in other traps. Though this difference in the tuning parameter for classical and quantum fluctuations stays within the crossover interval $\Delta T_X$ or $\Delta n_X$. The convex nature of $P_d$ vs $n$ as opposed to the concave nature of $P_d$ vs $T$ basically follows from the similar nature of the loss of six coordinated neighborhood.

It will be erroneous, however, to conclude from Fig.~\ref{fig6} that the crossover mechanism has a similarity for the thermal and quantum melting. For description of its critical properties a $d$-dimensional quantum system can be mapped onto the corresponding $(d+1)$-dimensional effective classical system. A prominent example is the quantum Ising model. The size of the extra dimension is determined by the inverse temperature $\beta=1/k_B T$ and diverges as $T \rightarrow 0$. Hence, at low temperatures quantum effects are always important. As a result, in a 2D macroscopic system the order of the phase transition is changed: second order for the thermal melting (KTHNY-type) is substituted by first order for the quantum melting (variation of the density parameter $r_s$ at fixed low $T$)~\cite{casula,Ast}. Detailed calculations done for the quantum 2D XY model provide the following scenario~\cite{Akopov}. The correlation function shows the crossover between the 2D and 3D behaviors at $R_c\sim 1/T$. For the distances $R \gtrsim R_c$ a system exhibits only the 2D character. Up to the distances $R< R_c$  there exists a quasi-long-range order and the correlation function shows the 3D behavior. Once the system size satisfies $L\sim R_c$ the 3d character dominates. By lowering the temperature, $R_c$ diverges and the 3D behavior can, in principle, be observed in an arbitrary system of finite size. 

\begin{figure}[t]
\includegraphics[width=9.5cm,height=12.0cm]{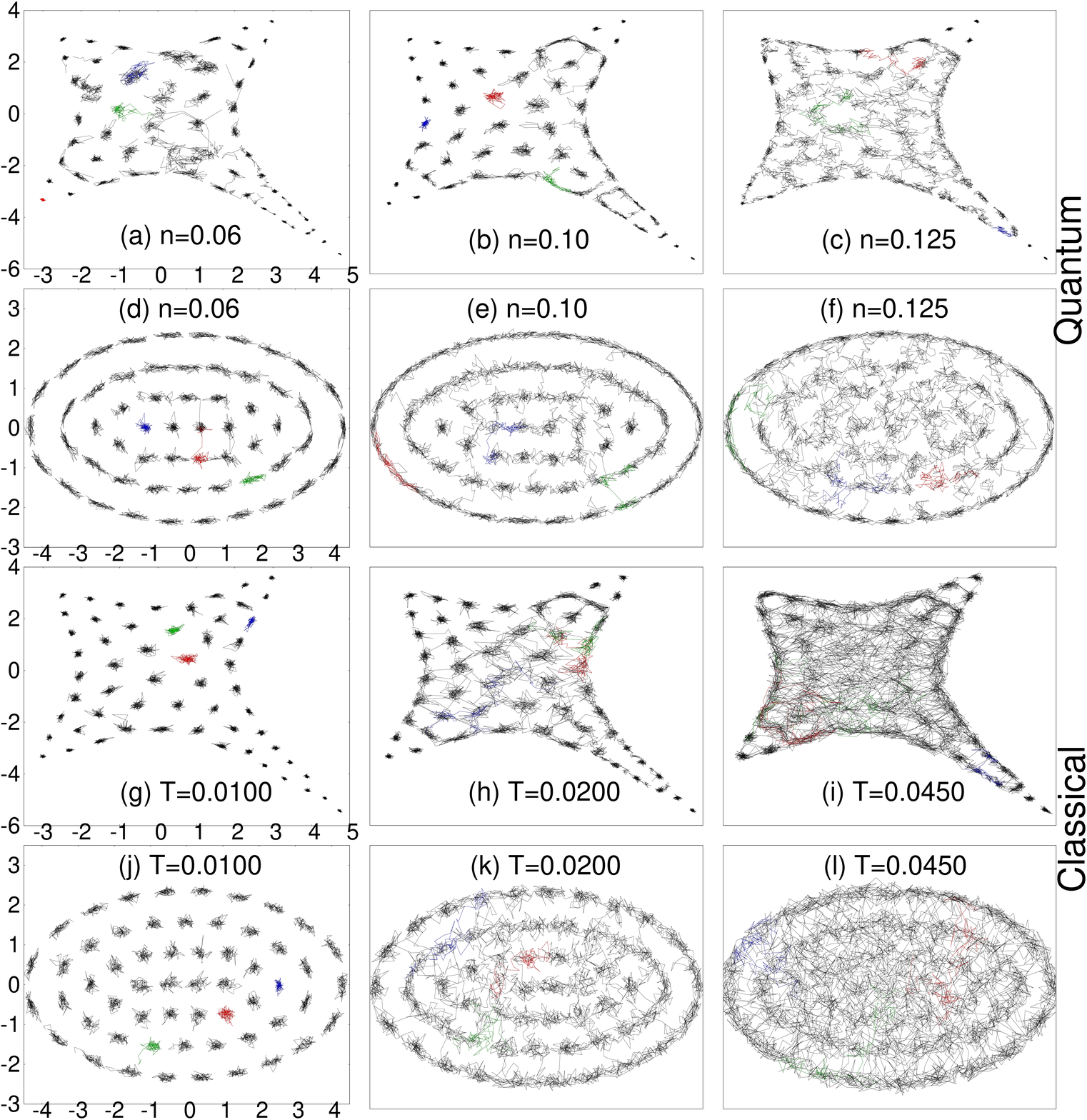}
\caption{
(color online) Particle trajectories are shown in different confinements to contrast the mechanism of quantum and classical melting. The top two rows show the trajectory of the BC-coordinates of the particles in irregular and elliptic confinements for three different $n$ values, whereas, the bottom two rows show the same during classical melting at three $T$. 
In the classical IWM, melting starts by coherent movement of certain particles along a tortuous path while other particles rattle around their equilibrium locations. In the quantum counterpart, we observe harmonic motion of all particles around their equilibrium position brings in the quantum melting.
Colored lines show how the individual particles diffuse into the system. At lowest $T$ and $n$ particles fluctuate mostly around their mean positions. With increasing $T$ or $n$ the particle diffusion grows.}
\label{fig7}
\end{figure}

We can conclude, that the applicability of the KTHNY physics in the phase diagram, cf. Fig.~\ref{fig4}c, is limited to the region of thermal melting. In the quantum domain, by increasing the system size, we expect to find a first order phase transition. Its identification lies in a sharp increase of the number of defects,
as found for circular and irregular confinement (admittedly, a similar sharpness is missing for the irregular geometry). In contrast, if a transition is of the KTHNY-type, the defects are increasing continuously. In our case, this crucial distinction is masked by the solid-liquid crossover due to finite size effects. In the quantum case (which in the thermodynamic limit would correspond to a first order transition) we observe a more rapid increase of the defect fraction compared to the thermal melting, cf. Fig.~\ref{fig6}.

To further elaborate possible differences in the melting mechanism, we present the snapshots of particle trajectories in the irregular and elliptical traps during the quantum and classical crossover, cf. Fig.~\ref{fig7}. In the quantum case, only the BC-particle coordinates~(\ref{bc}) are studied to make comparison on equal grounds. It is evident from Fig.~\ref{fig7} that the quantum melting sets in primarily by the harmonic fluctuations around equilibrium positions, and the mean squared displacements are increased with $n$. In contrast, the thermal melting commences by connecting neighboring particles over long distances with coherent displacements making a long and tortuous paths of collective motion in certain regions of the system. Such a collective motion across the crossover also occurs for the circular and elliptic traps, however, the paths are less rambling and follow the underlying confinement symmetry.

\subsection{Distribution of ${\cal L}$ with classical and quantum fluctuations}

Our results from the previous section indicated that the quantum crossover from the 'solid' to 'liquid' occurs via harmonic fluctuations, whereas the corresponding classical `melting' begins by aligning displacements of certain particles over long distances while the other particles remain confined around their equilibrium positions. We expect to illustrate this key difference using the distribution of Lindemann parameter obtained from the statistics over all particles. This is because a harmonic melting produces a Gaussian distribution of (positive) ${\cal L}$ centered at zero, but our picture of thermal melting would produce its broad distribution with additional peaks (corresponding to the drifting particles along the tortuous path of delocalization) and with no obvious symmetry. With this motivation, we 
calculate ${\cal L}_i$ separately for each particle $i$ ($i=1$ to $N$) by averaging it over all the MC configurations. We then represent their probability density of displacements as $P({\cal L})$ by constructing (normalized) histograms with these ${\cal L}_i$.
For our calculations, ${\cal L}_i$ is defined as ${\cal L}_i = \sqrt{\langle|u_i|\rangle}$ for $i$th particle, and $u_i = {\bf r}_i -{\bf r}^{(0)}_i$. Here ${\bf r}^{(0)}_i$ is the position of the $i$th particle in the initial configuration (after equilibration).
For a justified comparison of the results from quantum crossover with the corresponding classical ones, the quantum version of ${\cal L}_i$ is evaluated using the BC-coordinates, see Eq.~4.

\begin{figure}[t]
\includegraphics[width=9.5cm,height=11.0cm]{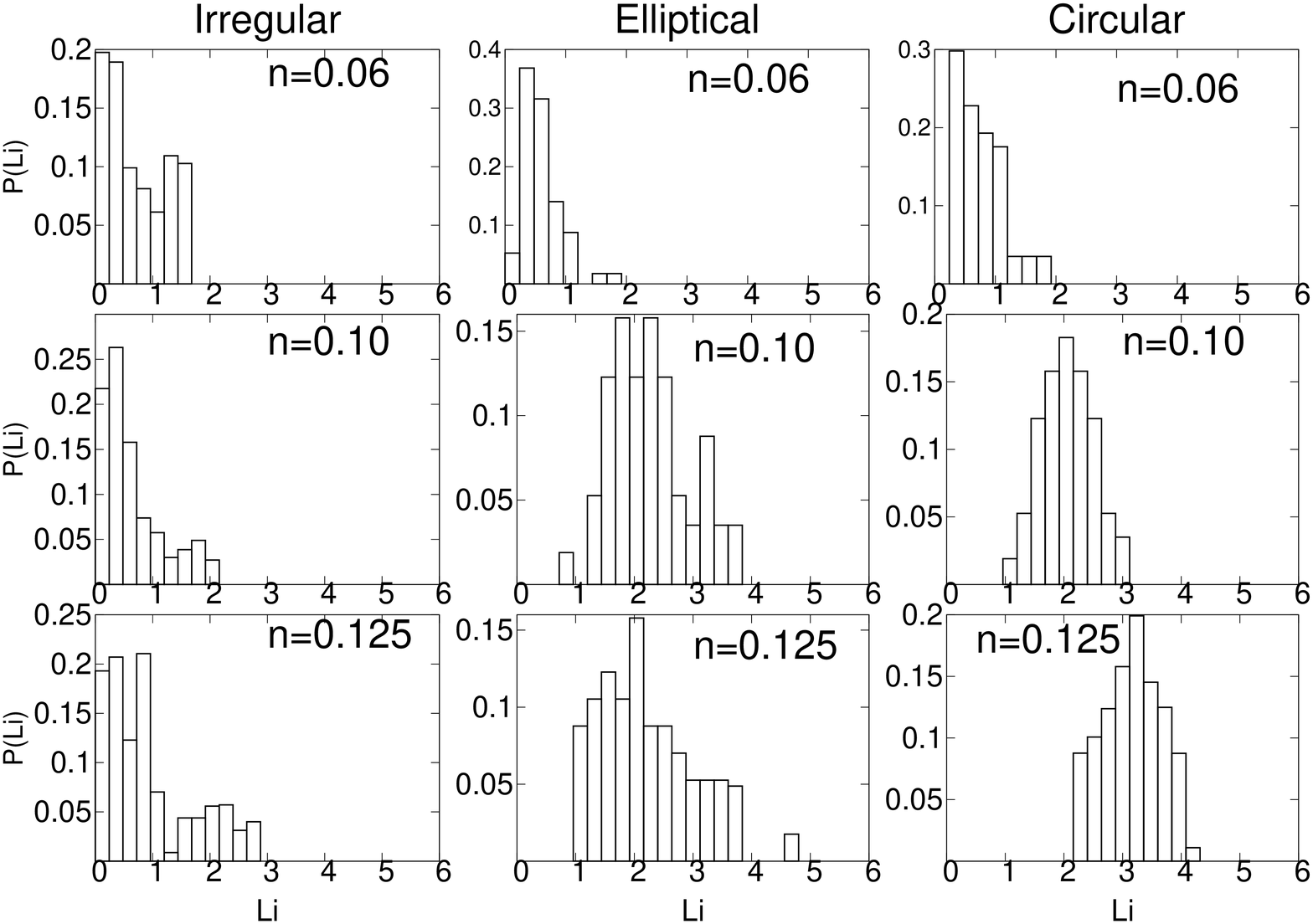}
\caption{
Evolution $P({\cal L}^{BC})$ vs $n$ at $T=0.005$ is presented where BC is used to calculate the distribution to compare it with the classical case presented in Fig.~\ref{fig9}. In circular case, the distribution is narrowly peaked at some high values of $n$, while in the irregular case, $P({\cal L}^{BC})$ is rather broad starting from very low values of ${\cal L}^{BC}$.
}
\label{fig8}
\end{figure}

\begin{figure}[t]
\includegraphics[width=9.5cm,height=11.0cm]{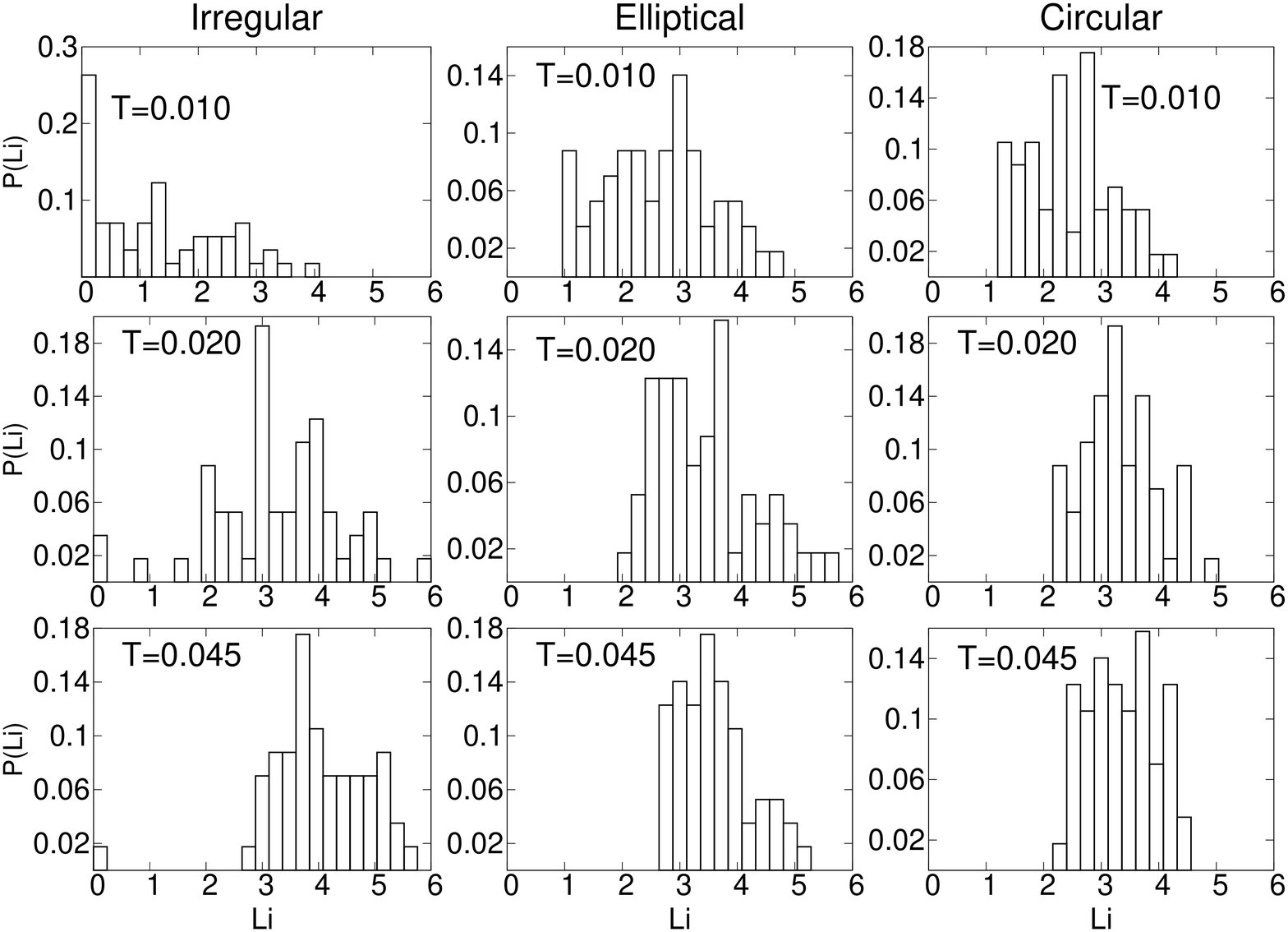}
\caption{
The evolution of classical $P({\cal L})$ with increasing thermal fluctuations in the three traps. At large $T=0.045$, beyond thermal crossover the distribution is narrow and symmetric in the circular case, while in the irregular and elliptical case, it has a long-tail like structure. The influence of underlying geometry of the trapping potential can be seen from this criteria.
}
\label{fig9}
\end{figure}

Our results for the evolution of $P({\cal L}^{BC})$ for all the three traps are presented in Fig.~\ref{fig8}. In the absence of a good resolution of our data, we focus on the the important qualitative features of $P({\cal L}^{BC})$ leaving out their details (note that a total of $N=57$ particles is not enough for drawing quantitative conclusions). Considering first the irregular trap (left panels), we notice that the distribution has a sharp peak at ${\cal L}^{BC}\approx 0$ for small $n$ (cf. $n=0.06$) representing the quantum `solid'. The peak broadens yielding a longer tail with increasing $n$, however, still a significant fraction of particles contribute to ${\cal L}^{BC}\approx 0$ even beyond the crossover, cf. $n=0.125 > n_X$ (note, that for the irregular trap $n_X \approx 0.10(1)$, cf. Fig.~2d). While most particles tend to diffuse in the quantum `liquid' phase, the extent of diffusion is not the same for all particles. Particularly the ones near the corners and long limbs of the irregular trap, keep `rattling' only around their mean positions contributing weight near ${\cal L}^{BC}\approx 0$.
 The structure and evolution of $P({\cal L}^{BC})$ for irregular trap is qualitatively consistent with our picture of ``melting via harmonic fluctuations".

The elliptical and circular traps, on the other hand, possess spatial symmetry that results in spatial shells of population of particles along the symmetry direction~\cite{BedPeet94, Filinov01}. Thus, for the circular and elliptic traps $P({\cal L}^{BC})$ provides also a measure of the angular shell rotation. Detailed analysis of our data demonstrates that the shift of the centroid of $P({\cal L}^{BC})$ with $n$ from ${\cal L}^{BC} \approx 0$ to a finite value, leaving no weight to $P({\cal L}^{BC}\approx 0)$, for circular and elliptic traps is due to the loss of the angular order of shells. On the other hand, the shape of $P({\cal L}^{BC})$ and its width is contributed by the inter-shell hops of the particles. The absence of spatial shells for the irregular confinement makes the evolution of $P({\cal L}^{BC})$ qualitatively different from those in the traps with spatial symmetries.

The evolution of $P({\cal L})$ in Fig.~9 with $T$ in case of thermal melting is similar to Fig.~8 insofar as the broad features are concerned (particularly for the elliptic and circular traps), though important differences persist. For example, $P({\cal L})$ is already broad for the irregular trap at low $T=0.01$. An increase of $T$ produces larger fraction of particles whose ${\cal L}$ are distributed over a wide range. Such evolution is in sharp contrast with the ``harmonic melting" of irregular quantum `solid' discussed above. Nevertheless, we close this section noting that it is the evolution of $P({\cal L})$ that points towards differences in the `melting' coming from the geometry of the confinement, which is not discerned from most other observables. 

\section{Conclusion}
In conclusion, we have studied the structural order-disorder transition for classical and quantum IWM of trapped Coulomb-interacting Boltzmann particles.
Confinements with or without spatial symmetries leave signatures somewhat differently on the
crossover from {\em Wigner} molecule to liquid-like phase.
However, the quantum zero point motion seems to screen the boundary effects in all traps, which results in a broadly similar evolution of observables across the crossover defining different melting criteria.

We have introduced several independent melting criteria which for different symmetries of the trap point towards a unique density interval of the crossover for the quantum melting, i.e. $[n_X-\Delta n,n_X+\Delta n]$ with $n_X\approx 0.11$ being the mid-point and $\Delta n\approx 0.01$. This interval via Eq.~(\ref{rs}) can be translated into the range of bulk densities: $46 \lesssim r_s \lesssim 69$. The critical value, $r_s \approx 66.5$, found for the quantum melting of a 2D Coulomb bulk system of Boltzmannons~\cite{casula} is close to the upper bound of the crossover region. A lower crossover scale for a trap than the bulk critical value is expected~\cite{GhosalNat07}, because the loss of translational symmetry in traps reduces the ability for the particles to delocalize and the resulting `solid' survives up to a smaller $r_s$.

Our detailed analyses of melting in the irregular confinement allowed us to reconstruct a precise phase diagram which shows several phases: a) thermal quasi-classical liquid, b) {\em Wigner} solid and c) quantum liquid. Here we note that the properties of the liquid phase strongly depend on the quantum statistics. This will modify the absolute values of $(n_X, \Delta n)$ characterizing the crossover. However, our discussion of the melting criteria introduced for Boltzmann particles will remain valid also for fermionic and bosonic IWM.

Of course, boltzmannons are just a model case to capture quantum diffraction effects. This neglects quantum statistics effects of fermions or bosons. However, in the crystal phase where particles are strongly localized, these effects are expected to be small~\cite{HuangBook87}.  In particular, the suppression of the  fermi liquid behavior by the strong  interactions is well-known in the $^3$He system~\cite{Glyde}. The statistics effects become more pronounced near the solid-liquid crossover line. For $^4$He systems long exchanges of indistinguishable bosons play an important role in the stabilizing the superfluid phase~\cite{Ceperley}. In contrast, Fermi statistics is known to stabilize the solid phase which is directly reflected in the different critical $r_s$-values of crystallization in 2D bulk: $r_s\approx 37$ for fermions~\cite{Tanatar89} and $r_s\approx 60$ for bosons~\cite{Palo}. As a consequence, the statistics has an effect on the precise location of the crossover line, and should definitely influence a location of the crossover region in finite systems~\cite{our_rs}. At the same time, we expect that the role played by quantum statistics will be reduced due to finite size and geometry effects.

It is indeed reassuring that the phase boundary produced by independent criteria match reasonably well within the statistical errors. We have uncovered a new feature in the phase diagram for the irregular trap in the form of the {\em re-entrant} solidification or the weakening of liquidity upon increasing temperature in the quantum regime. Identification of such a phenomenon would be an intriguing experimental proposition. 

The quantum crossover is associated with a sharply increasing defect fraction (disclination) in the region around $n_X$. 
The sharpness in the proliferation of defects is weaker in the irregular trap, however, the presence of spatial symmetries in case of circular or elliptic confinements strengthens such sharpness hinting that the crossover turns into a first order transition in the thermodynamic limit. Nevertheless, our extrapolation based on a small system can hardly differentiate between a weakly first order and nearly second order transitions. They pose serious numerical challenges even for a bulk system. In contrast, the corresponding classical transition is governed by the KTHNY physics. This suggests that the quantum melting mechanism should be quite similar to the one in a harmonic solid in $3$D. We hope that our work will contribute to
 pivotal understanding of the melting in disordered $2$D materials and will stimulate a new body of experimental research.

\section{Acknowledgement}

DB acknowledges the hospitality of the Physics Department of Kiel University where part of this work was performed.
This work is supported by the Deutsche Forschungsgemeinschaft via project FI 1252/2 and SFB-TR24 and by Ministry of Human Resource Development (MHRD), Govt of India.


\begin{thebibliography}{99}
\bibitem{our_rs} It is interesting to note that our $r_s$-values are located in between the critical values known for 2D bulk Fermi and Bose systems~\cite{Tanatar89,Palo}.
\end{thebibliography}
\end{document}